\documentclass{aa}  

\usepackage{graphicx}
\usepackage{txfonts}
\usepackage{booktabs}
\usepackage{microtype}
\usepackage[modulo,switch]{lineno}
\usepackage{natbib}
\usepackage{appendix}
\usepackage{footnote}
\usepackage[usenames]{xcolor}
\begin{document}

   \title{Gravity mode offset and properties of the evanescent  zone in red-giant stars}

 %  \subtitle{I. Overviewing the $\kappa$-mechanism}

   \author{S. Hekker \inst{1,2} \and Y. Elsworth \inst{3,2} \and G.C. Angelou \inst{1,2}}

   \institute{Max-Planck-Institut for Solar System Research, Justus-von-Liebig-Weg 3, D-37077 G\"ottingen, Germany\\
              \email{Hekker@mps.mpg.de}
         \and Stellar Astrophysics Centre, Department of Physics and Astronomy, Aarhus University, Ny Munkegade 120, DK-8000 Aarhus C, Denmark
         \and School of Physics and Astronomy, University of Birmingham, Birmingham B15 2TT, UK }

   \date{Received September 15, 1996; accepted March 16, 1997}

% \abstract{}{}{}{}{} 
% 5 {} token are mandatory
 
  \abstract
   {The wealth of asteroseismic data for red-giant stars and the precision with which these data have been observed over the last decade calls for investigations to further understand the internal structures of these stars.}
   {The aim of this work is to validate a method to measure the underlying period spacing, coupling term and mode offset of pure gravity modes that are present in the deep interiors of red-giant stars. We subsequently investigate the physical conditions of the evanescent zone between the gravity mode cavity and the pressure mode cavity.}
   {We implement an alternative mathematical description, compared to what is used in the literature, to analyse observational data and to extract the underlying physical parameters that determine the frequencies of mixed modes. This description takes the radial order of the modes explicitly into account, which reduces its sensitivity to aliases. Additionally, and for the first time, this method allows us to constrain the gravity mode offset $\epsilon_{\rm g}$ for red-giant stars.}
   {We find that this alternative mathematical description allows us to determine the period spacing $\Delta\Pi$ and the coupling term $q$ for the dipole modes within a few percent of literature values. Additionally, we find that $\epsilon_{\rm g}$ varies on a star by star basis and should not be kept fixed in the analysis. Furthermore, we find that the coupling factor is logarithmically related to the physical width of the evanescent region normalised by the radius at which the evanescent zone is located. Finally, the local density contrast at the edge of the core of red giant branch models shows a tentative correlation with the offset $\epsilon_{\rm g}$.}
   {We are continuing to exploit the full potential of the mixed modes to investigate the internal structures of red-giant stars; in this case we focus on the evanescent zone. It remains, however, important to perform comparisons between observations and models with great care as the methods employed are sensitive to the range of input frequencies.}


   \maketitle
%
%________________________________________________________________
\section{Introduction}

The long near-uninterrupted high-precision photometric timeseries data from the CoRoT and \textit{Kepler} space missions now allow for the investigation of the internal structures of stars. One of these structure features in red-giant stars is the evanescent zone between the cavity in which oscillations are present with buoyancy as restoring force (g-mode cavity) and the cavity in which pressure is the restoring force (p-mode cavity). The location, shape and width of this evanescent zone may all play a role in the coupling between these cavities and the characteristics of the observed dipole modes (modes with degree $l=1$), which have a mixed pressure-gravity nature \citep{takata2016a,takata2016b,mosser2017}. We note here that generally the dipole modes rather than quadrupole modes (modes with degree $l=2$) are used to investigate the interior conditions in stars. This is due to the fact that 1) the coupling is weaker at higher degrees leading to very small amplitudes of the modes with a significant g component and 2) the spacing between the mixed components reduces as a function of degree making quadrupole mixed modes and their period spacings harder to resolve. In the remainder of the paper we discus dipole modes, in all cases where no degree is indicated.

The underlying characteristics of the gravity part of the mixed modes are the asymptotic period spacing ($\Delta\Pi$), the coupling factor ($q$) and an offset ($\epsilon_{\rm g}$). The asymptotic period spacing is the period spacing between pure gravity modes (g modes) in the limit where the degree of the mode is much lower than the radial order $n$ (i.e. $l << n $). The spacings in period between individual mixed modes is in theory always smaller than the asymptotic value due to the coupling with a pressure mode. The coupling factor provides insight into the strength of the coupling between the g-mode cavity and the p-mode cavity, with $q=0$ for no coupling and $q=1$ indicating maximum coupling. The parameter $\epsilon_{\rm g}$ is a phase term accounting for the behaviour near the turning points of modes \citep[e.g.][]{hekker2016}.

Various approaches have been employed to determine the parameters of the mixed modes. For a subset of known red giants, \citet{bedding2011} derived the most prominent period spacing by taking the
power spectrum of the power spectrum for dipole modes, where the mode frequencies were expressed in period and the amplitude of the power spectrum was set to zero in regions not containing $l=1$ modes. Using this period spacing they presented period-\'echelle diagrams.  In such diagrams  frequency ($\nu$) is shown as a function of period ($\Pi$) modulo period spacing ($\Delta\Pi$), see e.g. panel E of Fig.~\ref{KIC010123207}. The frequencies of the mixed modes in consecutive acoustic radial orders are stacked on top of each other and show a typical "S-shape". The value of $\epsilon_{\rm g}$ determines the absolute position of the "S-shape" pattern in the period-\'{e}chelle diagram, while $q$ determines the steepness of the central segments; a shallow transition in the case of strong coupling and a steep transition in the case of weak coupling. 

\citet{mosser2012} presented the asymptotic expansion for the frequencies ($\nu$) of mixed modes based on \citet{unno1989}:
\begin{equation}
\nu=\nu_{n_p,l}+\frac{\Delta\nu}{\pi}\arctan\left[q\tan\pi\left(\frac{1}{\Delta\Pi_l\nu}-\epsilon_{\rm g}\right)\right],
\label{eq:M12}
\end{equation}
where $\nu_{n_p,l}$ is the frequency of the pressure mode (with radial order $n_p$ and degree $l$) with which the gravity modes are coupled, $\Delta\Pi_l$ is the asymptotic period spacing of modes with degree $l$, and $\Delta\nu$ is the large frequency separation between modes of the same degree and consecutive radial orders of acoustic modes. One commonly adopted assumption is to take $\epsilon_{\rm g}$ as a fixed value, either zero or one half depending on the definitions used. By fixing $\epsilon_{\rm g}$ it is possible to determine the period spacing and coupling strength in an iterative manner.  This formulation has successfully been applied in many cases \citep[e.g.,][]{mosser2014,buysschaert2016}.  
In fact, \citet{buysschaert2016} were the first to leave $\epsilon_{\rm g}$ as a free parameter in Eq.~\ref{eq:M12} and concluded that this enables a more robust analysis of both the asymptotic period spacing and the coupling factor. However, their method left $\epsilon_{\rm g}$ ill-defined with a large confidence interval \citep[e.g.][]{buysschaert2016}.

\citet{datta2015} developed an automated way to find the optimal value of $\Delta\Pi_l$ by measuring the alignment of the "S-shape" and the symmetry in the period-\'echelle diagram, i.e. implicitly assuming a fixed value for $\epsilon_{\rm g}$. Furthermore, to extract reliable period spacings they used a Monte Carlo approach, i.e., period spacings were computed for 10\,000 realisations of the data, for which the frequencies were randomly perturbed within their uncertainties. The distribution of the $\Delta\Pi_l$ results of each perturbation could then be used to compute a value with uncertainties for each discrete solution of $\Delta\Pi_l$ as well as the probability of the solution. Hence, in this analysis multiple results of the period spacing for a particular star with their probabilities were presented.

Following this, \citet{mosser2015} realised that the observed and asymptotic period spacings can be related through the ratio ($\zeta$) between the kinetic energy in the radiative cavity and the total kinetic energy \citep[e.g.][]{goupil2013,deheuvels2015}. This relation allowed \citet{mosser2015} to use $\zeta$ to compute stretched periods of mixed modes, where the modes in the stretched period-\'echelle diagram line up along vertical ridges. \citet{vrard2016} used this concept to develop an automated tool to compute gravity period spacings for over 6\,100 red giants observed with the \textit{Kepler} telescope. 
This method using $\zeta$ is particularly powerful as it requires only an approximate determination of $\Delta\nu$, $\Delta\Pi_1$, and the frequency position of the dipole pressure modes. Moreover, this method is applied to the full power spectrum and does not require knowledge of frequencies of individual modes.

\citet{mosser2017} subsequently investigated the coupling factors, i.e. $q$, of thousands of red giants with the intent to provide physical constraints on the regions surrounding the radiative core and the hydrogen-burning shell. They found that weak coupling is  present in only the most evolved stars on the red-giant branch. Larger coupling factors are measured at the transition between subgiants and giants as well as in core helium burning (CHeB) stars. 

An alternative mathematical description that is consistent with Eq.~\ref{eq:M12} has been proposed by \citet{jcd2012}, and developed further by \citet{jiang2014} \citet{cunha2015} and \citet{hekker2016} to compute the theoretical frequencies of mixed modes. This formalism has so far been used to compute frequencies of mixed dipole modes for models. Here, we investigate the performance of this formalism in determining $\Delta\Pi$, $q$ and $\epsilon_{\rm g}$ when applied to observed data. This method explicitly uses frequencies as well as the value of the radial order of mixed modes and has the advantage that it can be applied to models (with frequencies computed in an independent way) as well as to observational data. This allows us to investigate 1)  how close the period spacings computed from individual mixed-mode frequencies and from the integral of the Brunt-V\"ais\"al\"a frequency are; 2) the physical conditions of the evanescent zone connected with the coupling term. Additionally, we comment on the physical meaning of $\epsilon_{\rm g}$ and what the impact is of choosing a different set of frequencies to derive the period spacing.

\section{Method}
The formalism proposed by \citet{jcd2012} is as follows:
\begin{equation}
\Pi_{n\,l}= \frac{1}{\nu_{n\,l}}=\Delta \Pi_l \left [|n|+ \epsilon_{\rm g}+ \frac{1}{2} -\frac{\Phi(\nu_{n\,l})}{\pi} \right ],
\label{eq:mixfreq}
\end{equation}
where $|n|$ is the absolute value of the numerical radial mode order (see Section 2.1). 
Additionally, $\Phi(\nu_{n\,l})$ satisfies 
\begin{equation}
\tan \Phi(\nu_{n\,l}) = q \cot \left(\pi \left ( \frac{\nu_{n\,l}}{\Delta\nu}-\epsilon_{p\,l} \right) \right),
\label{eq:phi}
\end{equation}
where we assume that $\Delta\nu$ has the same value as obtained from a linear fit through the radial modes, i.e. a typical way to extract it from the frequency spectra, and $\epsilon_{p\,l}$ is an offset for the acoustic modes of degree $l$ (see for more details Section 5.4).

To apply the formalism outlined here to observed dipole frequencies we start by supplying the algorithm with an initial guess of $\Delta\Pi_l$. We then compute the radial orders of the oscillation modes as per Eq.~\ref{eq:n} and apply a $\chi^2$ fit procedure to Eq.~\ref{eq:mixfreq}. Here, we allow $|n|$ to vary by an integer and keep $\Delta\nu$ fixed to find the values of $\Delta\Pi$, $q$, $\epsilon_{\rm g}$ and $\epsilon_{\rm p\,1}$ that give a best fit to the observed frequencies of modes with a particular degree ($l$) and azimuthal order ($m$).
This is based on the fact that for slowly rotating red giants we have computed the frequency of the $m=0$ component of modes with $m\neq0$ using the description by \citet{mosser2012rot}. To obtain a best fit with a lowest $\chi^2$ we use a grid of initial guesses of $\Delta\Pi$ ranging for red giant branch stars from 50-100 seconds in steps of 0.01 seconds and for low-mass core helium burning stars from 170 - 360 seconds in steps of 0.1 seconds. In this way we obtain for each initial $\Delta\Pi$ the radial order of the modes and a computed value for $\Delta\Pi, q, \epsilon_{\rm g}$ and $\epsilon_{p\,1}$ as well as a measure of the goodness of fit through the $\chi^2$ value. We note here that we have defined $\epsilon_{p\,1}$ to have a value between 0.5 and 1.5 similar to $\epsilon_{p\,0}$.

In this goodness of fit we have to account for the fact that we expect a larger number of gravity modes ($N_{\nu_g}$) in a $\Delta\nu$ interval for lower values of $\Delta\Pi$ \citep{mosser2012}:
\begin{equation}
N_{\nu_g} \cong \frac{\Delta\nu}{ \Delta\Pi \, \nu_{\rm max}^{2}},
\label{nr}
\end{equation}
where $\nu_{\rm max}$ is the frequency of maximum oscillation power. 
%Due to this effect the $\chi^2$ value will increase as a function of $\Delta\Pi$. Here, we aim to find a minimum with respect to this increasing underlying trend. The amount of increase of $\chi^2$ as a function of $\Delta\Pi$ depends on both the number of expected and observed oscillation modes and can vary on a star by star basis. Therefore, we mitigated this effect empirically by normalising the $\chi^2$ values. To this end, we compute a median smooth with a width of 10 seconds for the $\chi^2$ values obtained from all fits with different values of the initial guess $\Delta\Pi$ as a function of the computed $\Delta\Pi$. We divide the $\chi^2$ values by this smooth, which is equivalent to high-pass filtering the $\chi^2$ of our full set of results. Due to this normalisation the values of the $\chi^2$ only have a meaning in a relative sense. 

To incorporate uncertainties in the observed frequencies as well as correlations between different parameters we use a Monte Carlo approach with 100 iterations to perturb the observed frequencies randomly within their uncertainties. We apply the described method for each set of perturbed frequencies. %Per iteration we store the results with the lowest 5\% of $\chi^2$ and later select the solutions with the lowest $\chi^2$ per iteration. From these lowest $\chi^2$ solutions we cluster the results with the same $|n|$ in case of multimodal solutions and compute the median $\Delta\Pi$, $q$, $\epsilon_{\rm g}$ and $\epsilon_{\rm p\,1}$ with the 16\% and 84\% quartiles as the uncertainties. We do this for all clusters and report the most probable cluster. The cluster probability is defined by the ratio of the number of results in the cluster to the $\chi^2$ values of these results, i.e. both the number of results increase the probability as well as a low $\chi^2$.

We note here that this formalism does not take into account effects of glitches, i.e. sudden internal structure changes visible as variations in the oscillation frequencies and thus also period spacings \citep[described in detail by][in the case of buoyancy glitches]{cunha2015}.  As stated by \citet{cunha2015}, buoyancy glitch-induced variations occur on the red-giant branch only at the luminosity bump, and after the red-giant branch only in the early phases of helium core burning and at the beginning of helium shell burning. Hence, we expect that for many stars it is not necessary to perform a glitch analysis in order to extract period spacings. 

\subsection{Radial order}
Fundamental to the formalism discussed here is knowledge of the radial order of every dipole feature that is present. 
Following the asymptotic analysis, the period of a pure gravity mode can be expressed as
\begin{equation}
\Pi_{n\,1}= \Delta\Pi(|n_{\rm g}| + \epsilon_{\rm g} +1/2).
\end{equation}
From this we can estimate the absolute value of the gravity mode order $n_{\rm g}$ as
\begin{equation}
|n_{\rm g}| \cong \frac{1}{\Delta\Pi \, \nu_{n\,1}},
\end{equation}
where we neglect the $\epsilon_{\rm g} +1/2$ as $|n_{\rm g}| >> \epsilon_{\rm g} +1/2$. 
The final estimate of the radial order $n$ of a specific frequency of a mixed mode is a combination of the pure gravity radial order and the pure pressure radial order ($n_{\rm p}$) and can be computed as :
\begin{equation}
n = n_{\rm g} + n_{\rm p} \cong  \frac{-1}{\Delta\Pi  \, \nu_{n\,1}} + \left(\frac{\nu_{n\,1}}{ \Delta\nu} - \epsilon_{p\,1}\right),
\label{eq:n}
\end{equation}
where we used the general convention that gravity mode orders are indicated with negative values. We note that $\Delta\nu$ and $\epsilon_{p\,0}$ are computed from a linear fit through the radial frequencies and that a first estimate of $\epsilon_{p\,1}$ is obtained using the correction for the degree according to $\epsilon_{p\,1} \approx \epsilon_{p\,0} + 1/2$ (see also Section 5.4).
Combined with the requirement that the radial order of each mode should be an integer and that $n$ should increase with frequency, we can compute $n$ provided that all other parameters are known. To obtain the radial mode orders of modes with $m \neq 0$ we estimate the rotational splitting using a Lorentzian profile as proposed by \citet{mosser2012rot} to identify the frequency of the underlying unsplit mode and use that frequency to compute the radial order in the same way as outlined above.

\section{Data}
In this section we detail the data to which we apply the method outlined above. These data comprise theoretical models and observational data of both red giant branch (RGB) stars and core helium burning (CHeB) stars. 

\begin{table*}
        \centering
        \begin{minipage}{\linewidth}
        \caption{Core helium burning models from \citet{constantino2015} used in the current work. The first three columns give the figure number, colour code used in the figure and the period spacing computed by \citet{constantino2015}. The last columns provide the identification we use in this work, a comment concerning the regularity of the behaviour of the observed period spacing ($\Delta P$) with frequency, and a rough value of our determined $\Delta\Pi$ for comparison purposes. The * indicates models for which we reliably recover $\Delta\Pi$ and that are used in the further analysis in this paper.}
        \label{tab:chebmods}
        \renewcommand{\footnoterule}{}
        \centering
        \begin{tabular}{rllllc} % four columns, alignment for each
                \hline
Fig.~\# & colour & $\Delta\Pi$ & our work & regularity $\Delta P$ & $\Delta\Pi_{\mathrm{this\,work}}$\\
\hline
9 & black & 240~s & CHeBmodel 0 & irregular* & $\sim$241~s\\
9 & blue & 238~s & CHeBmodel 1 & irregular* & $\sim$243~s\\
10 & black & 247~s & CHeBmodel 2 & irregular & --\\
10 & blue & 247~s& CHeBmodel 3 & regular* & $\sim$249~s\\
12 & black & 252~s & CHeBmodel 4 & irregular* & $\sim$253~s\\
12 & orange & 253~s & CHeBmodel 5 & regular* & $\sim$253~s\\
12 & blue & 253~s & CHeBmodel 6 & regular* & $\sim$253~s\\
13 & blue & 253~s & CHeBmodel 7 & regular* & $\sim$253~s\\
13 & magenta & 314~s & CHeBmodel 8 & regular* & $\sim$316~s\\
14 & black & 278~s & CHeBmodel 9 & irregular & --\\
14 & cyan & 281~s\footnote{If the calculation includes only the region exterior to the chemical discontinuity then $\Delta\Pi$ = 315~s \citep{constantino2015}.} & CHeBmodel 10 & regular/spiky & --\\
16 & black & 273~s & CHeBmodel 11 & semi-regular* & $\sim$274~s \\
16 & orange & 271~s & CHeBmodel 12 & regular/spiky & --\\
16 & cyan & 264~s & CHeBmodel 13 & regular/spiky & -- \\
17 & orange & 268~s & CHeBmodel 14 & semi-regular* & $\sim$277~s\\
                \hline
        \end{tabular}
        \end{minipage}
\end{table*}

\subsection{Red giant branch stars}
We used the three red giant branch models described by \citet{datta2015} to test the application of the formalism discussed in this manuscript. The models by \citet{datta2015} are 1~M$_{\odot}$ models at different stages of hydrogen shell burning computed using the MESA stellar evolution code \citep{paxton2011}. These models were chosen because they are computed independent of the development of the formalism discussed here and $\Delta\Pi$ has been provided. 
From the models we prepared sets of frequencies that mimic "observational" data by selecting modes in three frequency ranges with 5, 7 and 9 radial orders centred around $\nu_{\rm max}$. In this frequency range we kept either all frequencies, or we selected modes with normalised inertias (with respect to radial mode inertias) such that on average we have 5 modes per (acoustic) radial order. In the analysis we neglected the fact that we know the radial orders of these oscillation modes. We assumed an uncertainty of 0.008~$\mu$Hz on all dipole frequencies. This value is approximately the frequency resolution of the  $\sim$~4-year long timeseries of \textit{Kepler} data. We show throughout the paper the results obtained for the mode sets with all dipole modes in a 5$\Delta\nu$ wide frequency range.

Additionally, we applied our method to frequencies from stars observed by \textit{Kepler}. We used frequencies for the sample of stars presented by \citet{datta2015} and \citet{corsaro2015}\footnote{We increased the uncertainties on the frequencies as provided by \citet{corsaro2015} by a factor of three, as it was shown that these were underestimated by roughly that factor (Corsaro et al. erratum in preparation).}, as well as KIC~4447888 \citep{dimauro2016}. Two of the three stars analysed by \citet{datta2015} are part of the sample analysed by \citet{corsaro2015}. We used both sets of data as the frequency values have been determined independently. In total 21 stars were treated, with $\Delta\Pi$ values ranging from 68.5 to 90~s (see also Fig.~\ref{freqs} for a visual representation of the distribution of dipole frequencies).%In Fig.~\ref{freqs} we show the dipole frequencies of all RGB stars that we used for the analysis. We discuss this figure further in Section~4 together with the results.

\subsection{Core Helium burning stars}
In the last few years several studies \citep[e.g.][]{bossini2015,constantino2015,lagarde2016,bossini2017} have investigated the physics that needs to be included to remedy the discrepancy between $\Delta\Pi$ values of CHeB stars derived from observations \citep[e.g.][]{mosser2014} and from models with standard physics included.
In this work we use the models described by \citet{constantino2015} who computed 1~M$_{\odot}$ solar metallicity CHeB models using the MONSTAR stellar evolution code \citep{lattanzio1986,campbell2008,constantino2014}. These authors apply different core-mixing schemes at different phases of core Helium burning, i.e. just after the helium flash (or non-degenerate onset of He-core burning) all the way towards exhaustion of helium in the core. We applied our method to the models in Figs~9, 10, 12, 13, 14, 16 and 17 of \citet{constantino2015}, which were provided by the authors. Table~\ref{tab:chebmods} provides some basic information about the models. We use `regular', `irregular', `regular/spiky' and `semi-regular' to classify the observed period spacing ($\Delta P$) of the models as a function of frequency. This classification is  determined from a visual inspection of the $\Delta P$ versus frequency figures presented by \citet{constantino2015}. We classify a star as regular when $\Delta P$ (relatively) smoothly approaches minima at the pressure dominated dipole modes and maxima at the radial modes \citep[see for an example the orange curve in the top panel of Fig.~12 of][]{constantino2015}. When such dips are not present as is the case in the top panel of Fig.~9 of \citet{constantino2015} we assign the classification `irregular'. With `semi-regular' we indicate models for which $\Delta P$ shows dips as a function of frequency, but with a significant amount of irregular structure on top of that. We call dips that are very narrow such as the cyan curves in Figs~14 and 16 `regular/spiky'. %As our method depends on regular behaviour of the observed period spacing $\Delta P$, we don't expect it to work for the models with irregular behaviour. Nevertheless, we applied it to all models to verify this. 
We applied our method to all models that we have at our disposal.

The "observational" data were obtained from the models in the same way as per the RGB models, described in the previous subsection. Additionally, we applied the procedure to an Li-rich star (KIC~5000307) in the red clump \citep{silvaaguirre2014}. 

\begin{figure*}
\centering
\includegraphics[width=0.9\linewidth]{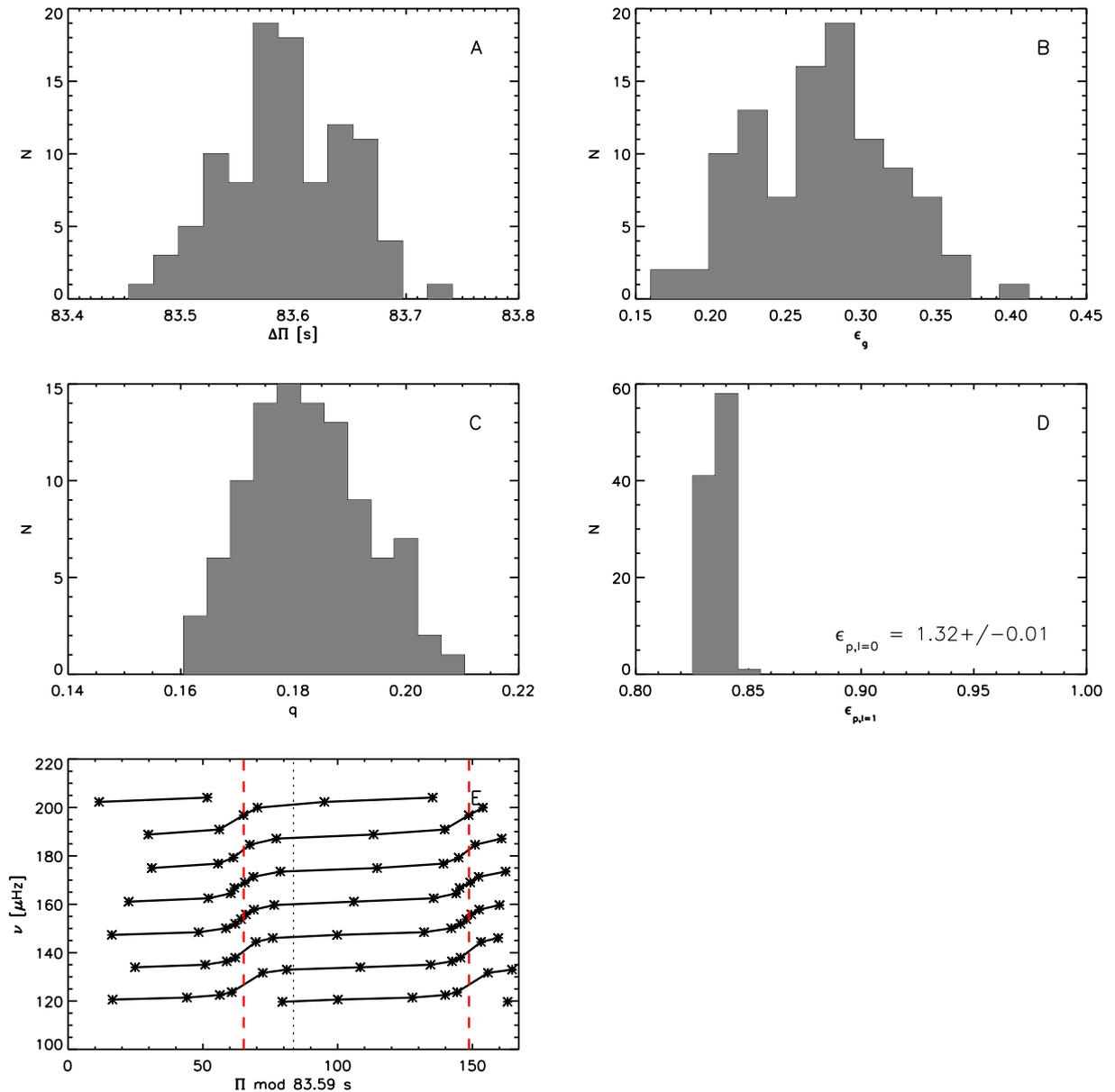}
\caption{Results for KIC~10123207 using the frequencies from \citet{corsaro2015}. Panel A-D: histograms of $\Delta\Pi$, $\epsilon_{\rm g}$, $q$, and $\epsilon_{\rm p\,1}$ (see Introduction for the meaning of these parameters). The $\epsilon_{\rm p\,0}$ value determine from radial modes is indicated in the legend of panel D. A period-\'{e}chelle diagram using the $\Delta\Pi$ obtained in this work is shown in panel E. The vertical red dashed lines indicate $((\epsilon_{\rm g} +0.5)$~mod~1)$*\Delta\Pi$, i.e. the position of the most g-dominated modes according to the fitted values. Note that the period-\'echelle diagram is shown twice separated by the dotted vertical line.}
\label{KIC010123207}
\end{figure*}

\begin{figure*}
\centering
\includegraphics[width=0.9\linewidth]{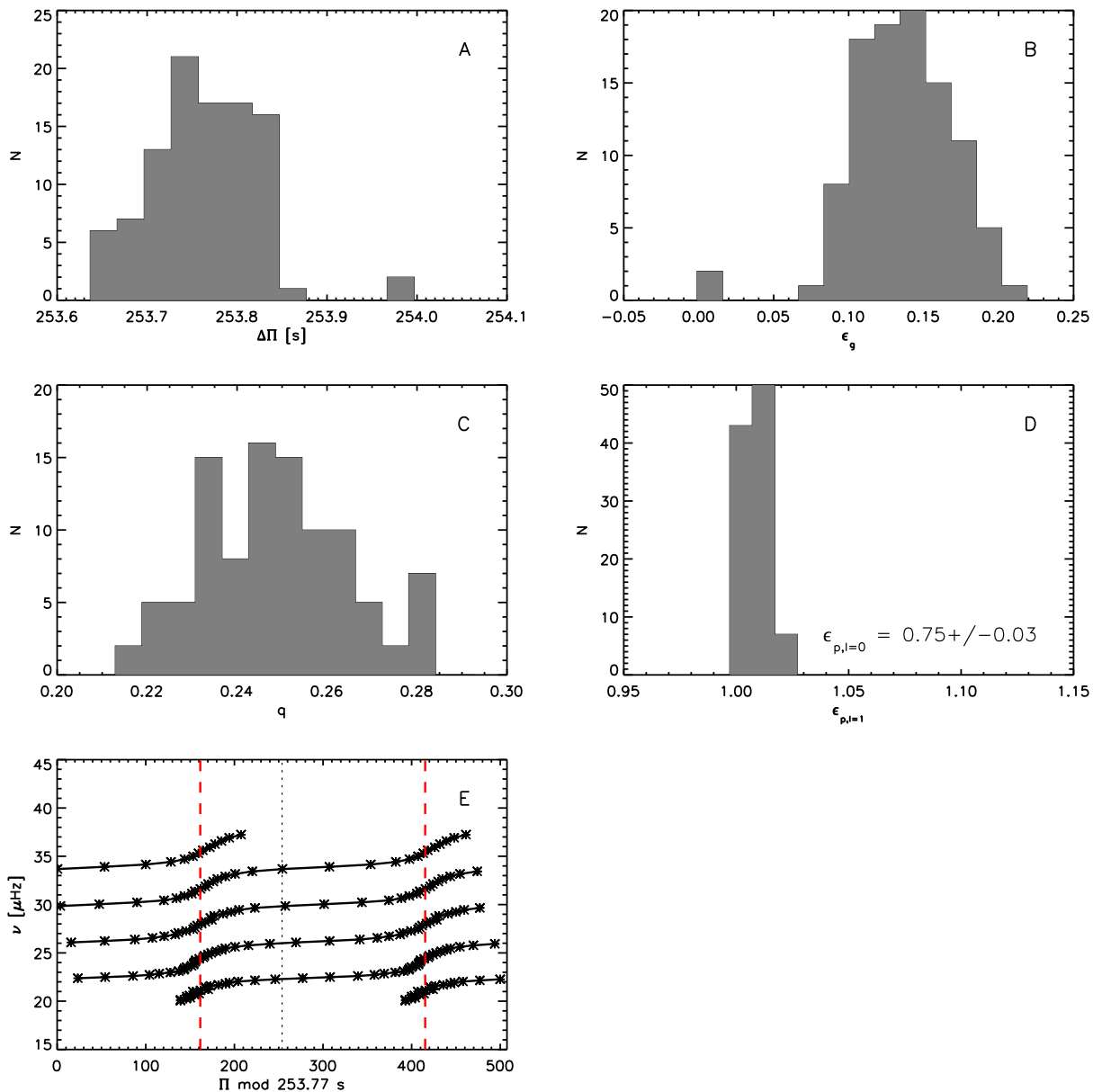}
\caption{Same as Fig.~\ref{KIC010123207} for CHeBmodel 7 with all frequencies selected in a $5\Delta\nu$ wide range.}
\label{CHeBmodel7}
\end{figure*}

\begin{table*}
\caption{Results for the RGB stars and models. For most stars the results are based on the frequencies presented by \citet{corsaro2015}. Stars indicated with superscript `a' show results based on frequencies presented by \citet{datta2015} and the star indicated with superscript `b' shows results based in frequencies presented by \citet{dimauro2016}. For the models we show results for the mode set with all dipole frequencies included in a $5\Delta\nu$ wide range.}
\label{resultstabRGB}
\centering
\begin{tabular}{lccccccc}
\toprule
star & $\Delta\Pi_{\rm lit}$ [s] & $\Delta\Pi$ [s] & $\epsilon_{\rm g}$ & $q$ & $\epsilon_{\rm p\, 1}$ & $\Delta\nu$ [$\mu$Hz] & $p$\\
\hline
KIC003744043 &  75.98 &76.05$^{0.09}_{0.10}$ & 0.9$^{0.1}_{0.1}$ & 0.140$^{0.010}_{0.010}$ & 0.810$^{0.010}_{0.047}$ & 9.841$\pm0.006$ & 1.00
\\
\hline
KIC004448777$^{\rm b}$ &  89.87 &89.33$^{0.02}_{0.03}$ & 0.29$^{0.01}_{0.01}$ & 0.155$^{0.002}_{0.001}$ & 0.8841$^{0.0006}_{0.0007}$ & 16.921$\pm0.006$ & 0.66
\\
\hline
KIC005866737$^{\rm a}$ &  68.49 &66.014$^{0.005}_{0.006}$ & 0.008$^{0.012}_{0.008}$ & 0.093$^{0.006}_{0.005}$ & 0.686$^{0.003}_{0.003}$ & 6.499$\pm0.006$ & 0.42
\\
\hline
KIC006117517 &  76.91 &76.86$^{0.06}_{0.06}$ & 0.09$^{0.10}_{0.08}$ & 0.15$^{0.02}_{0.03}$ & 0.872$^{0.005}_{0.005}$ & 10.031$\pm0.006$ & 1.00
\\
\hline
KIC006144777 &  79.23 &79.04$^{0.02}_{0.04}$ & 0.24$^{0.04}_{0.03}$ & 0.121$^{0.007}_{0.007}$ & 0.80$^{0.01}_{0.01}$ & 10.956$\pm0.004$ & 1.00
\\
\hline
KIC007060732 &  77.10 &77.76$^{0.05}_{0.04}$ & 0.14$^{0.05}_{0.06}$ & 0.129$^{0.006}_{0.008}$ & 0.799$^{0.003}_{0.004}$ & 10.853$\pm0.004$ & 0.81
\\
\hline
KIC007619745 &  79.17 &79.04$^{0.06}_{0.09}$ & 0.13$^{0.09}_{0.05}$ & 0.144$^{0.009}_{0.011}$ & 0.869$^{0.006}_{0.051}$ & 13.059$\pm0.006$ & 0.95
\\
\hline
KIC008366239 &  86.77 &87.84$^{0.15}_{0.09}$ & 0.25$^{0.07}_{0.10}$ & 0.14$^{0.01}_{0.01}$ & 0.866$^{0.005}_{0.004}$ & 13.619$\pm0.006$ & 0.97
\\
\hline
KIC008475025 &  74.80 &74.46$^{0.03}_{0.06}$ & 1.00$^{0.09}_{0.05}$ & 0.126$^{0.009}_{0.007}$ & 0.764$^{0.004}_{0.003}$ & 9.572$\pm0.004$ & 1.00
\\
\hline
KIC008718745 &  79.45 &79.99$^{0.03}_{0.03}$ & 0.34$^{0.05}_{0.04}$ & 0.146$^{0.008}_{0.006}$ & 0.788$^{0.003}_{0.003}$ & 11.363$\pm0.005$ & 0.98
\\
\hline
KIC009145955$^{\rm a}$ &  76.98 &77.023$^{0.008}_{0.008}$ & 0.983$^{0.010}_{0.011}$ & 0.155$^{0.002}_{0.002}$ & 0.8800$^{0.0009}_{0.0009}$ & 10.882$\pm0.005$ & 1.00
\\
\hline
KIC009145955 &  77.01 &76.78$^{0.06}_{0.05}$ & 0.31$^{0.08}_{0.06}$ & 0.17$^{0.01}_{0.02}$ & 0.827$^{0.007}_{0.005}$ & 10.941$\pm0.005$ & 1.00
\\
\hline
KIC009267654 &  78.41 &78.13$^{0.09}_{0.09}$ & 0.93$^{0.13}_{0.10}$ & 0.14$^{0.01}_{0.01}$ & 0.798$^{0.007}_{0.007}$ & 10.239$\pm0.004$ & 0.79
\\
\hline
KIC009475697 &  75.70 &75.54$^{0.04}_{0.05}$ & 0.23$^{0.09}_{0.07}$ & 0.18$^{0.02}_{0.02}$ & 0.797$^{0.005}_{0.005}$ & 9.806$\pm0.004$ & 0.56
\\
\hline
KIC009882316 &  80.59 &80.42$^{0.08}_{0.10}$ & 0.14$^{0.09}_{0.07}$ & 0.19$^{0.01}_{0.01}$ & 0.860$^{0.008}_{0.007}$ & 13.602$\pm0.007$ & 1.00
\\
\hline
KIC010123207 &  83.88 &83.59$^{0.06}_{0.06}$ & 0.28$^{0.05}_{0.06}$ & 0.18$^{0.01}_{0.01}$ & 0.836$^{0.004}_{0.006}$ & 13.629$\pm0.007$ & 1.00
\\
\hline
KIC010200377$^{\rm a}$ &  81.54 &81.300$^{0.010}_{0.009}$ & 0.338$^{0.009}_{0.011}$ & 0.155$^{0.002}_{0.002}$ & 0.7820$^{0.0006}_{0.0008}$ & 12.501$\pm0.007$ & 1.00
\\
\hline
KIC010200377 &  81.58 &81.46$^{0.04}_{0.03}$ & 0.15$^{0.04}_{0.05}$ & 0.19$^{0.02}_{0.01}$ & 0.895$^{0.005}_{0.005}$ & 12.377$\pm0.004$ & 1.00
\\
\hline
KIC010257278 &  79.81 &79.72$^{0.07}_{0.05}$ & 0.08$^{0.07}_{0.05}$ & 0.143$^{0.009}_{0.009}$ & 0.81$^{0.06}_{0.01}$ & 12.114$\pm0.005$ & 0.89
\\
\hline
KIC011353313 &  76.00 &77.15$^{0.09}_{0.12}$ & 0.0$^{0.2}_{0.1}$ & 0.15$^{0.01}_{0.02}$ & 0.772$^{0.017}_{0.010}$ & 10.724$\pm0.006$ & 1.00
\\
\hline
KIC011913545 &  77.84 &77.79$^{0.09}_{0.06}$ & 0.08$^{0.07}_{0.08}$ & 0.123$^{0.014}_{0.008}$ & 0.790$^{0.036}_{0.002}$ & 10.092$\pm0.004$ & 0.98
\\
\hline
KIC011968334 &  78.10 &77.79$^{0.07}_{0.05}$ & 0.45$^{0.07}_{0.09}$ & 0.13$^{0.01}_{0.01}$ & 0.822$^{0.004}_{0.004}$ & 11.363$\pm0.005$ & 1.00
\\
\hline
KIC012008916 &  80.47 &81.4$^{0.3}_{0.2}$ & 0.2$^{0.2}_{0.2}$ & 0.09$^{0.01}_{0.02}$ & 0.804$^{0.044}_{0.008}$ & 12.834$\pm0.005$ & 0.62
\\
\hline
RGBmodel 0 &  82.61 &82.239$^{0.007}_{0.008}$ & 0.317$^{0.009}_{0.009}$ & 0.141$^{0.002}_{0.002}$ & 0.5212$^{0.0008}_{0.0008}$ & 11.977$\pm0.003$ & 1.00
\\
\hline
RGBmodel 1 &  73.49 &73.216$^{0.007}_{0.006}$ & 0.30$^{0.01}_{0.02}$ & 0.126$^{0.005}_{0.004}$ & 0.501$^{0.002}_{0.002}$ & 7.136$\pm0.003$ & 1.00
\\
\hline
RGBmodel 2 &  62.15 &62.036$^{0.005}_{0.006}$ & 0.14$^{0.05}_{0.04}$ & 0.079$^{0.009}_{0.008}$ & 1.357$^{0.005}_{0.004}$ & 4.186$\pm0.003$ & 0.55
\\
\bottomrule
\end{tabular}
\label{tab1}
\end{table*}

\begin{table*}
\caption{Same as for Table~\ref{tab1} for CHeB stars.}
\label{resultstabRC}
\centering
\begin{tabular}{lccccccc}
\toprule
star & $\Delta\Pi_{\rm lit}$ [s] & $\Delta\Pi$ [s] & $\epsilon_{\rm g}$ & $q$ & $\epsilon_{\rm p\,1}$ & $\Delta\nu$ [$\mu$Hz] & $p$\\
\hline
KIC005000307 & 319.95 &322.2$^{0.1}_{0.1}$ & 0.48$^{0.03}_{0.02}$ & 0.34$^{0.01}_{0.02}$ & 0.628$^{0.005}_{0.004}$ & 4.724$\pm0.002$ & 0.65
\\
\hline
CHeBmodel 0 & 240.00 & 241.98$^{0.01}_{0.02}$ & 0.989$^{0.006}_{0.006}$ & 0.24$^{0.02}_{0.02}$ & 1.051$^{0.006}_{0.004}$ & 3.527$\pm0.003$ & 0.86
\\
\hline
CHeBmodel 1 & 238.00 &243.32$^{0.06}_{0.07}$ & 0.36$^{0.05}_{0.03}$ & 0.22$^{0.02}_{0.01}$ & 1.049$^{0.004}_{0.006}$ & 3.584$\pm0.003$ & 0.99
\\
\hline
CHeBmodel 3 & 247.00 &249.83$^{0.04}_{0.07}$ & 0.57$^{0.04}_{0.03}$ & 0.23$^{0.02}_{0.01}$ & 0.941$^{0.006}_{0.005}$ & 3.611$\pm0.003$ & 1.00
\\
\hline
CHeBmodel 4 & 252.00 &253.30$^{0.04}_{0.05}$ & 0.15$^{0.03}_{0.03}$ & 0.249$^{0.009}_{0.014}$ & 1.004$^{0.006}_{0.004}$ & 3.734$\pm0.003$ & 0.68
\\
\hline
CHeBmodel 5 & 253.00 &253.80$^{0.06}_{0.04}$ & 0.31$^{0.02}_{0.03}$ & 0.26$^{0.02}_{0.01}$ & 1.400$^{0.004}_{0.005}$ & 3.908$\pm0.003$ & 1.00
\\
\hline
CHeBmodel 6 & 253.00 &253.82$^{0.06}_{0.04}$ & 0.33$^{0.02}_{0.03}$ & 0.26$^{0.02}_{0.01}$ & 1.400$^{0.006}_{0.005}$ & 3.908$\pm0.003$ & 1.00
\\
\hline
CHeBmodel 7 & 253.00 &253.76$^{0.06}_{0.06}$ & 0.14$^{0.03}_{0.03}$ & 0.25$^{0.02}_{0.02}$ & 1.008$^{0.006}_{0.005}$ & 3.740$\pm0.003$ & 0.98
\\
\hline
CHeBmodel 8 & 314.00 &315.96$^{0.07}_{0.08}$ & 0.29$^{0.03}_{0.02}$ & 0.29$^{0.02}_{0.01}$ & 1.349$^{0.006}_{0.004}$ & 3.810$\pm0.003$ & 1.00
\\
\hline
CHeBmodel 11 & 273.00 &274.09$^{0.06}_{0.06}$ & 0.43$^{0.02}_{0.03}$ & 0.28$^{0.02}_{0.01}$ & 1.334$^{0.004}_{0.004}$ & 4.090$\pm0.003$ & 1.00
\\
\hline
CHeBmodel 14 & 268.00 &277.30$^{0.06}_{0.07}$ & 0.49$^{0.03}_{0.02}$ & 0.28$^{0.01}_{0.02}$ & 1.325$^{0.004}_{0.006}$ & 3.743$\pm0.003$ & 1.00
\\
\bottomrule
\end{tabular}
\end{table*}

\section{Results}
In this section we present the results that we obtained with our method on the data described in the previous section. The results are presented in Tables~\ref{resultstabRGB} and \ref{resultstabRC}.

We show an illustration of the results for a RGB star in Fig.~\ref{KIC010123207}. In this figure, histograms of the results with lowest $\chi^2$ per Monte Carlo iteration as a function of $\Delta\Pi$, $\epsilon_{\rm g}$, $q$ and $\epsilon_{\rm p\,1}$ are shown in panels A $-$ D, respectively. Panel E shows a period-\'echelle diagram based on the derived $\Delta\Pi$ value (which value is quoted in the x-axis label). Panel E actually consists of a repeated \'echelle diagram to enable better visualisation the "S-shape". This is in principle possible for our solutions as we have $\epsilon_{\rm g}$ as a free parameter. We show the results for a CHeB star in a similar way in Fig.~\ref{CHeBmodel7}). 

We checked that the ratio of the uncertainty in $\epsilon_{\rm g}$ ($\sigma_{\epsilon_{\rm g}}$) to the relative uncertainty in $\Delta\Pi$ ($\sigma_{\Delta\Pi}/\Delta\Pi$) is roughly equal to $|n|$. This is generally satisfied for our results.

Below we discuss our results for both the red giant branch stars and core helium burning stars and compare them with literature values and$/$or values obtained from models. These comparisons are focussed on period spacings and coupling factors as these parameters are available in the literature or can be computed in an independent way from the models.

%\begin{figure}
%\centering
%\includegraphics[width=\linewidth]{reldPuncvsepsguncfinal2.eps}
%\caption{The ratio of the uncertainty in $\epsilon_{\rm g}$ ($\sigma_{\epsilon_{\rm g}}$) to the relative uncertainty in $\Delta\Pi$ ($\sigma_{\Delta\Pi}/\Delta\Pi$) as a function of absolute radial order $|n|$. The plus and minus uncertainties for observed stars are indicated with black and blue dots respectively, and in case of models with red and orange diamonds, respectively. The dotted line indicates the one-to-one relation.}
%\label{sigmavsn}
%\end{figure}

\begin{figure}
\centering
\includegraphics[width=\linewidth]{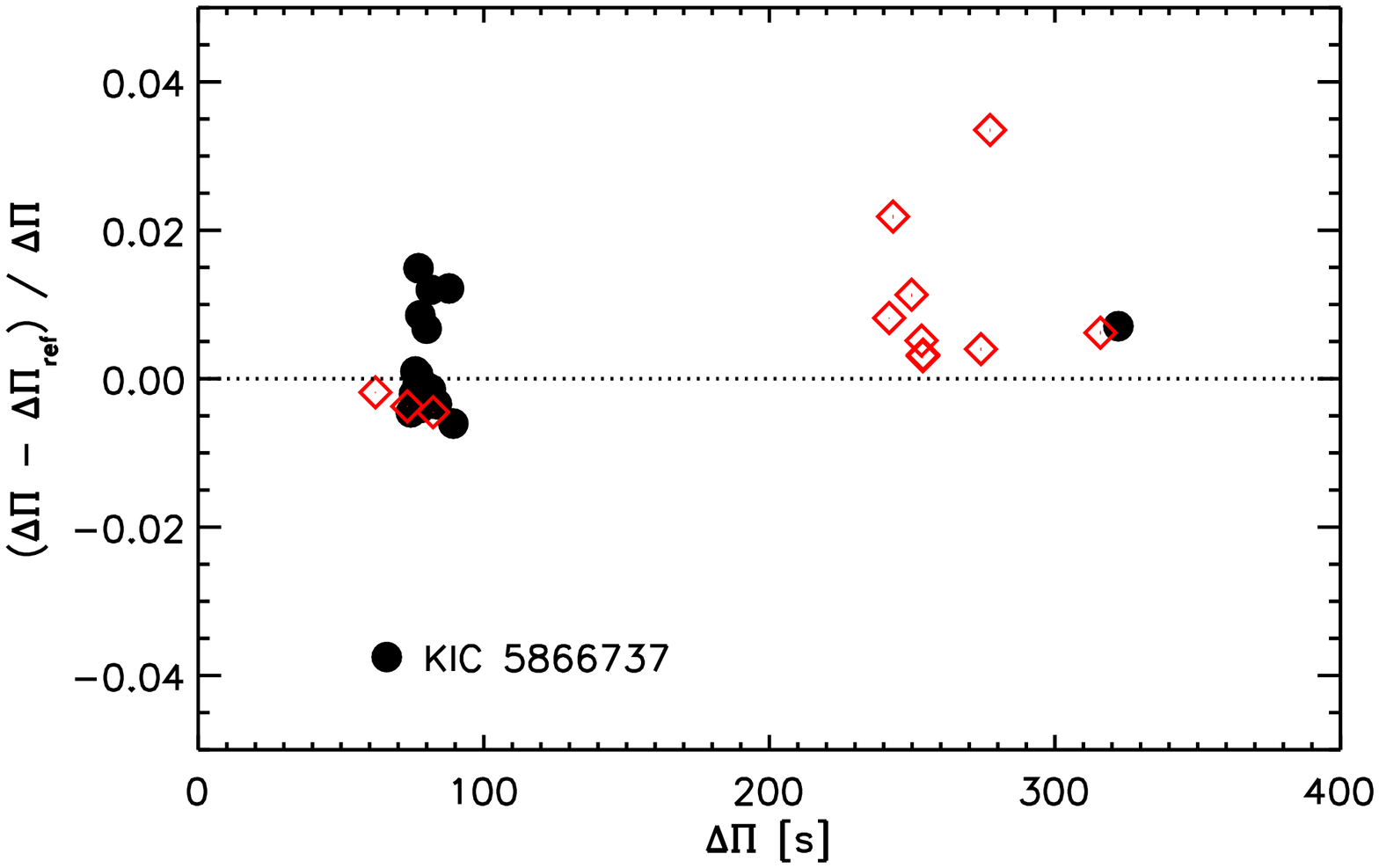}
\caption{Comparison of the period spacings derived in this work with reference values, where in case of the models the reference values are computed from the integral of the Brunt-V\"ais\"al\"a frequency (Eq.~\ref{eq:brunt}) and in case of real data reference values are observed values from the literature with updated values for KIC~6144777, KIC~7060732, see text for details]. Results for real stars and models are shown in black dots and open red diamonds, respectively.  Error bars are mostly smaller than the symbol size. Precise agreement is highlighted by the dotted line.}
\label{compreslit}
\end{figure}

\subsection{Red giant branch stars}

The $\Delta\Pi$ results obtained in this work for all three models described by \citet{datta2015} are in agreement with their values obtained from individual frequencies. The results are, however, most stable for model 0 and less so for the other more evolved models. This is due to the distribution of the dipole modes (see below) as well as the large absolute value of the radial order.

For 20 out of 21 observed stars (two stars, KIC009145955 and KIC010200377, are analysed twice with slightly different datasets) we find good agreement (better than 3\%) between the values of the period spacings obtained in this work and the results presented in the literature\footnote{We note that the results presented by \citet{corsaro2015}  are those obtained by \citet{mosser2012} and that for KIC~6144777, KIC~7060732 we have used updated values of $\Delta\Pi$~=~79.23~s and $\Delta\Pi$~=~77.10~s, respectively (Corsaro private communication).}. The relative differences are shown in Fig.~\ref{compreslit}. For one red giant branch star we find a somewhat larger discrepancy between the $\Delta\Pi$ value obtained in our work compared to the values obtained in the literature ($\sim$4\% difference): KIC~5866737. We discuss this star below in more detail. Additionally, we also compared our values with period spacings obtained by \citet{vrard2016} for the 16 stars that we have in common. The differences in the period spacings in this comparison are in all cases well within 1\%.

For KIC~5866737 we find that the ratio of the uncertainty in $\epsilon_{\rm g}$ ($\sigma_{\epsilon_{\rm g}}$) to the relative uncertainty in $\Delta\Pi$ ($\sigma_{\Delta\Pi}/\Delta\Pi$) is not roughly equal to $|n|$. At the same time, we find a $\Delta\Pi$ value that is roughly 4\% lower than obtained in the literature (see Fig.~\ref{compreslit}). KIC~5866737 is the most evolved star in our sample of observed stars, with dipole modes that are confined in a narrow range around the pressure dominated mode. The coupling is expected to decrease for more evolved stars along the RGB and we conclude that KIC~5866737 roughly indicates the limit along the RGB at which the method discussed here can produce reliable results in terms of gravity mode parameters. 
We note that for RGBmodel 1 ($\Delta\Pi \approx 72$~s) and RGBmodel 2 ($\Delta\Pi \approx 62$~s) the datasets that mimic observations also cause the method difficulties and fail in a number of cases.

\begin{figure}
\centering
\includegraphics[width=\linewidth]{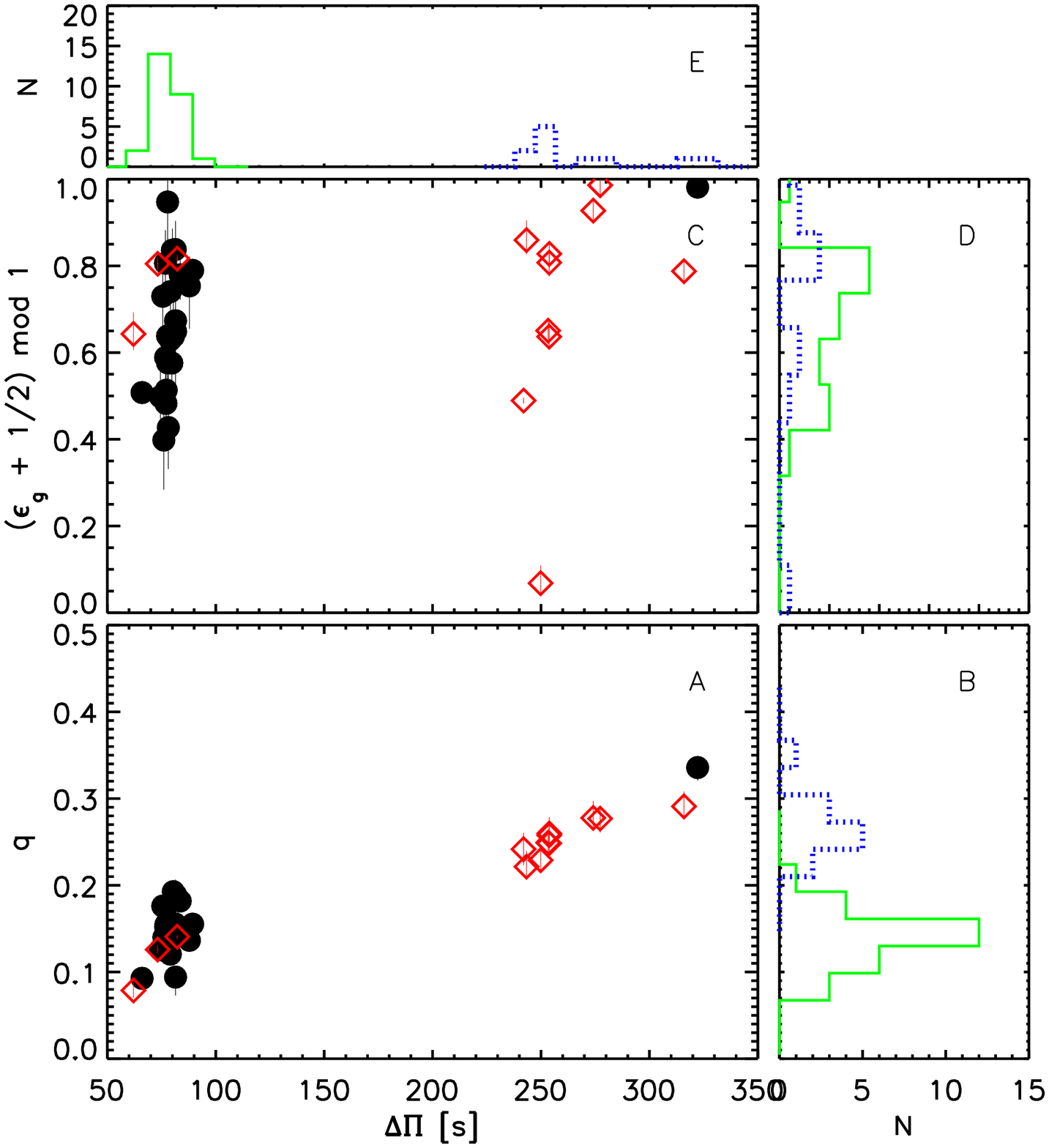}
\caption{Panel A: Coupling factor $q$ vs. $\Delta\Pi$ for all objects. Real stars are indicated with black dots and models with red diamonds. Panel B: histogram of $q$ for RGB stars (real stars + models) and CHeB stars in green solid and blue dotted lines, respectively. Panel C: same as panel A, but now for ($(\epsilon_{\rm g} +1/2)$~mod~1) vs. $\Delta\Pi$. Panel D: distribution of $(\epsilon_{\rm g}+1/2)$~mod~1. Panel E: the distribution of $\Delta\Pi$ results. The colour coding and linestyles in panel D and E are the same as in Panel B. In panels A and C uncertainties are over plotted. These are however in a number of cases smaller than the symbol size.}
\label{dPvsq_epsg}
\end{figure}

In Fig.~\ref{dPvsq_epsg} the coupling factor $q$ and offset $(\epsilon_{\rm g}+0.5)$~mod~1 are presented. We show here $(\epsilon_{\rm g}+0.5)$~mod~1 as this is the full offset that relates to the position of the g-dominated modes, i.e. the position of the "S-shape", in the period-\'echelle diagram. For the RGB stars ($\Delta\Pi < 100$~s, green histograms) we find $q$ values below 0.25 consistent with earlier results \citep{mosser2012,mosser2017}. For $(\epsilon_{\rm g}+0.5)$~mod~1 in RGB stars we find values between roughly 0.3 and 1. A discussion on this is presented in Section 5.

\begin{figure}
\centering
\includegraphics[width=\linewidth]{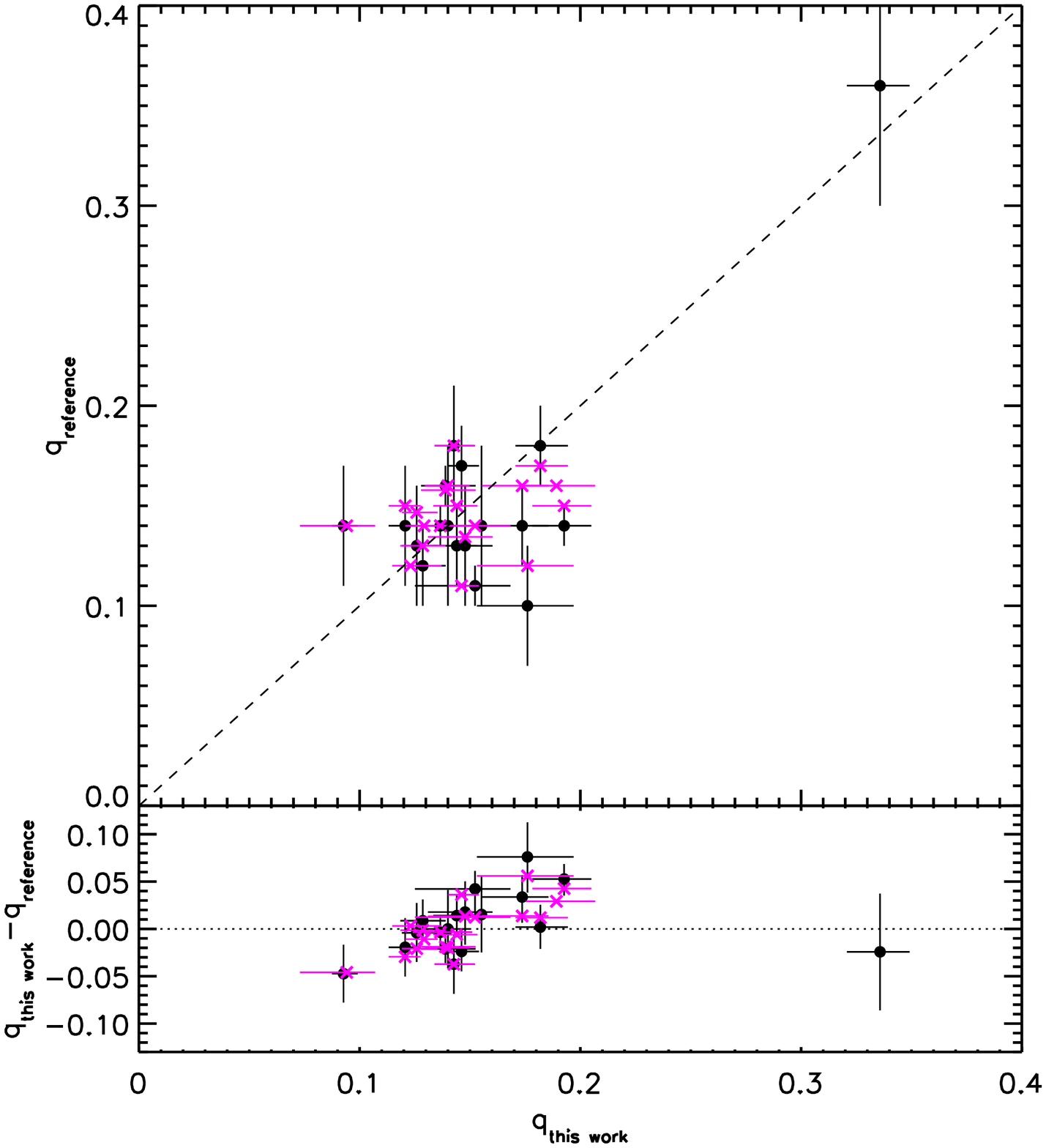}
\caption{Top: Comparison between the coupling factor $q_{\rm this~work}$ determined in this work and the values $q_{\rm reference}$ obtained by \citet[][black dots]{mosser2017} and Corsaro (private communication, magenta crosses) for the stars that we have in common. The dotted line indicates agreement. Bottom: the difference between the coupling in this work and in the reference values in the sense $q_{\rm this~work} - q_{\rm reference}$. Note that no uncertainties are provided by Corsaro.}
\label{qcomp}
\end{figure}

In Fig.~\ref{qcomp}, we compare our derived values for the coupling factor with the values obtained by \citet{mosser2017} and values from Corsaro (private communication) for the stars that we have in common. Generally, the values are consistent within their uncertainties (see top panel Fig.~\ref{qcomp}). However, we note that for RGB stars ($q < 0.25$) we find a linear correlation between the differences in $q$ (our values $-$ literature) vs. $q$ with a Pearson r coefficient of 0.7 (bottom panel of Fig.~\ref{qcomp}). We also computed the t-statistic and use a two-sided t-test to find that we can reject a relation with zero slope at $> 99$\% level. This correlation could  be related to the fact that in our analysis we have left $\epsilon_{\rm g}$ as a free parameter, while $\epsilon_{\rm g}$ was kept fixed in the analyses already present in the literature. % (see Section 5.3 for further comments).  

\subsection{Core Helium Burning stars}
In this work we analyse the models by \citet{constantino2015} as described in Table~\ref{tab:chebmods}. 
The results of these models are mostly summarised in Table~\ref{tab:chebmods}.
From these results we conclude that our method can be applied to CHeB stars with non-spiky behaviour, where some irregular behaviour can be accounted for. For CHeBmodel 2 we seem to be at the limit of the amount of irregularity the method can handle, although in some mode sets we do find the correct solution. For CHeBmodel 9, we find, in addition to the irregularity, spikes in the mode inertia profile that most likely hamper the determination of the period spacing. For the analysis in the remainder of this paper we take CHeB models 0,1, 3, 4, 5, 6, 7, 8, 11, 14 (indicated with a * in Table~\ref{tab:chebmods}) into account. The results for these models are also shown in Fig.~\ref{compreslit}.\newline
\newline
In terms of observations of CHeB stars we have frequencies for KIC~5000307, which is a lithium-rich star \citep{silvaaguirre2014}.  The current result for $\Delta\Pi$ is close ($\sim$1\% difference) to the value obtained by  \citet{silvaaguirre2014}. Such a value is consistent with the results by \citet{mosser2014} for a 1.5-2~M$_{\odot}$ CHeB star. 

For CHeB stars with $\Delta\Pi$ values that are deemed reliable, Fig.~\ref{dPvsq_epsg} shows the coupling factor $q$ and offset $\epsilon_{\rm g}$. For the CHeB stars ($\Delta\Pi > 200$~s, blue dashed histograms) we find $q$ values between 0.2 and 0.4 in agreement with earlier results \citep{mosser2012,mosser2017}. 

\section{Discussion}
We have shown that we can constrain $\Delta\Pi$, $q$, $\epsilon_{\rm g}$ and $\epsilon_{\rm p\,1}$  for stars with enough observed dipole modes using the formalism proposed by \citet{jcd2012}, \citet{jiang2014}, \citet{cunha2015} and \citet{hekker2016}. Based on these results, we first discuss what  the obtained parameters reveal about the internal structures of the stars and subsequently investigate the impact of different mode sets.

\subsection{Period spacing $\Delta\Pi$}
For the models we have compared the value of $\Delta\Pi$ obtained from frequencies using the approach outlined herein and the asymptotic value ($\Delta\Pi_{l,\rm asymptotic}$) computed through the integral of the Brunt-V\"ais\"al\"a frequency:
\begin{equation}
\Delta\Pi_{l,\rm asymptotic}=\frac{2\pi^2}{l(l+1)}\left(\int^{r_2}_{r_1}N\frac{dr}{r} \right)^{-1},
\label{eq:brunt}
\end{equation}
with $N$ the Brunt-V\"ais\"al\"a frequency and $r_1$ and $r_2$ the points the lower and upper turning points\footnote{In practice we computed one value for $\Delta\Pi_{l,\rm asymptotic}$ per model taking the integral over the total area of the Brunt-V\"ais\"al\"a frequency and not a separate value for each frequency with its specific turning points. The difference in the values is however negligible compared to the differences we discuss here.}.

The observed and computed asymptotic values are in broad agreement; for RGB models the value computed from the Brunt-V\"ais\"al\"a frequency are a bit higher (typically of the order 0.01 to a few times 0.1 second) than the one obtained from the frequencies. This is as expected from theory and in line with earlier results \citep[e.g.][]{datta2015}. For the CHeB models we find that our $\Delta\Pi$ values obtained from frequencies are larger than the asymptotic reference values (up to about 10 seconds, see Fig.~\ref{compreslit}). These relatively high values for $\Delta\Pi$ seem to be consistent with earlier findings by e.g.~\citet{mosser2014} and studied in more detail by  \citet{constantino2015}.  These authors report a systematic differences in $\Delta\Pi$ of CHeB stars between observations (using frequencies) and model predictions (using the Brunt-V\"ais\"al\"a frequency). The larger values of $\Delta\Pi$ obtained from frequencies may indicate that in CHeB stars the frequencies are not sensitive to the whole buoyancy cavity, possibly due to a discontinuity, or additional convective areas blocking the oscillations. \citet{bossini2017} indeed find that additional mixing in terms of core-overshooting can mitigate the differences in $\Delta\Pi$ between observations and models.

\begin{figure}
\centering
\includegraphics[width=\linewidth]{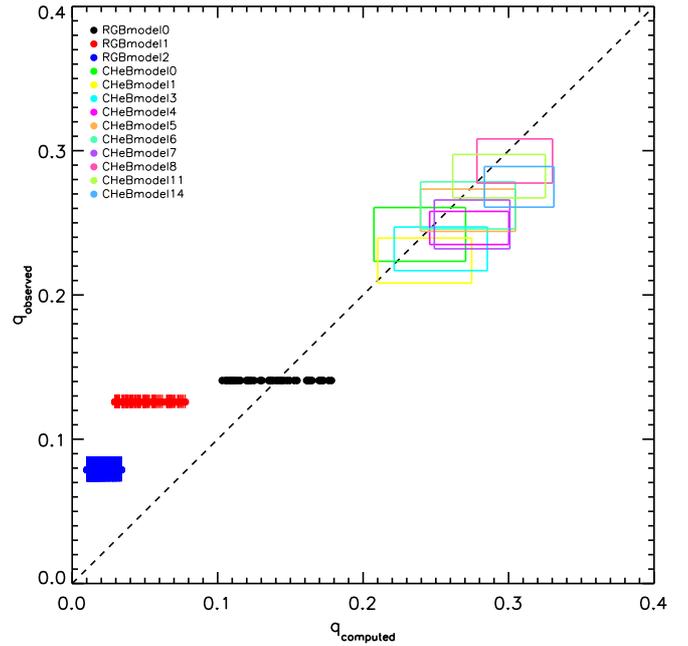}
\caption{The coupling factor computed using the method outlined in Section 2  ($q_{\rm observed}$) vs. the coupling factor for each frequency computed using Eqs~\ref{eq:q},~\ref{eq:T}~and~\ref{eq:kappa} ($q_{\rm computed}$). For visual purposes we show the individual points with uncertainties for the RGB models and for the CHeB models a rectangle that comprises the results including the uncertainties for that particular model. So the horizontal width of the box indicates the spread in $q$ computed through Eq.~\ref{eq:q} - \ref{eq:kappa} for all frequencies in a range of 5 times $\Delta\nu$ centred around $\nu_{\rm max}$. The legends show the colour with which each model is indicated.}
\label{qvsqobs}
\end{figure}

\begin{figure}
\centering
\includegraphics[width=\linewidth]{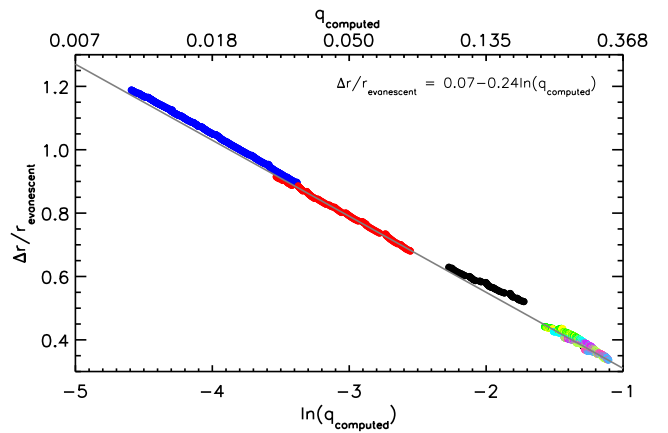}
\caption{The radial extent of the evanescent zone ($\Delta r$) normalised by the radius of the midpoint of the evanescent zone ($r_{\rm evanescent}$) versus the natural logarithm of the coupling factor ($\ln(q_{\rm computed})$) for each frequency in the frequency range $\nu_{\rm max} \pm 3.5\Delta\nu$ computed using Eqs~\ref{eq:q},~\ref{eq:T}~and~\ref{eq:kappa}. The colour coding is the same as in Fig.~\ref{qvsqobs}. The grey line indicates the fit as indicated in the legend (see text for more details). The values of $q_{\rm computed}$ are shown in the top axis.}
\label{qvsdrr}
\end{figure}

\subsection{Coupling term $q$}
It is known that $q$ provides information about the coupling of the wave in the gravity and acoustic cavity, with $q=0$ for no coupling and $q=1$ for full coupling \citep[e.g.][]{takata2016a}. In the current study, we would like to further our understanding as to the dependence of $q$ on physical parameters of the star.
Following \citet{takata2016a} and \citet{mosser2017} it is possible to compute $q$ for the models for each frequency using the following general formulation:
\begin{equation}
q=\frac{1-\sqrt{1-T^2}}{1+\sqrt{1-T^2}}
\label{eq:q}
\end{equation}
with $T$ the amplitude transmission coefficient, i.e. a measure of how much of the wave passes through a barrier (in this case the evanescent zone), computed as
\begin{equation}
T =  \exp \left (- \int_{D} \kappa~\mathrm{dr} \right )
\label{eq:T}
\end{equation}
with $D$ the radial extent of the evanescent region and $\kappa$ the radial wave vector:
\begin{equation}
\kappa=\frac{\sqrt{(S_1^2-\omega^2)(\omega^2-N^2)}}{c\omega}
\label{eq:kappa}
\end{equation}
with $S_1$ the Lamb frequency of a dipole mode, $N$ the Brunt-V\"ais\"al\"a frequency, $\omega$ the angular frequency equal to $2\pi\nu$ and $c$ the sound speed.

For the computed values of $q$ we find a different value for each oscillation mode due to the fact that each oscillation mode has a different $\kappa$ (Eq.~\ref{eq:kappa}) and encounters a slightly different evanescent zone as a consequence of the shape of the Brunt-V\"ais\"al\"a and Lamb frequencies. However, when extracting $q$ from observations, only one global value can be obtained as only the ensemble of dipole modes contains sufficient information to extract $q$. A comparison of the results of the computed $q$ values of the models with the value of $q$ obtained from the analysis using the method described in this paper is shown in Fig.~\ref{qvsqobs}. The results are generally consistent for $q>0.1$, while for weaker coupling we find that $q_{\rm observed}$ is over-estimated compared to the values obtained from the models. This is consistent with the results by \citet{mosser2017}, whom find that the computed coupling factor of a model high on the RGB is significantly smaller value than the observed value. The reason for this remains so far unclear.

We subsequently investigate the dependence of $q$ on physical parameters of the star and find empirically that $q$ shows a tight correlation with $\Delta r/r_{\rm evanescent}$, which is the radial extent of the evanescent zone normalised by the radius of the (midpoint of the) evanescent zone $r_{\rm evanescent}$. This correlation extends over the full range of $q$ that is covered by our models except for a small deviation for RGB model 0 at its highest $q$ values (see Fig.~\ref{qvsdrr}). We use symbolic regression to find a functional fit. %Symbolic regression is a type of optimisation that searches mathematical expressions to find a model that best fits the data. Thus rather than searching for the optimal parameters of a pre-defined expression (as is commonly done in regression techniques), the algorithm simultaneously optimises the functional form  and the parameters of the function. For the work presented here we used the software \textsc{Eureqa} \citep{schmidt09} distributed by Nutonian. This software package uses an evolutionary search to automatically perform symbolic regression. 
With this symbolic regression we find that $\Delta r/r_{\rm evanescent}$ is linearly related to $\ln (q_{\rm computed})$, where $q_{\rm computed}$ is computed using Eqs~\ref{eq:q}~-~\ref{eq:kappa}:
\begin{equation}
\Delta r / r_{\rm evanescent} =0.07-0.24\ln(q_{\rm computed}).
\label{qrel}
\end{equation}
This linear behaviour is expected theoretically (see Eqs~\ref{eq:q}-\ref{eq:kappa}) and vindicated by the use of the symbolic regression. We speculate that the deviation of RGB model 0 from the relation is because this model is earliest in evolution compared to the other models and still on its way to the homology that is apparent in the other models. We note that the coefficients in Eq.~\ref{qrel} are valid for 1~M$_{\odot}$ models with solar metallicity and may differ depending on mass and metallicity.

\subsection{Offset $\epsilon_{\rm g}$}

We now consider $\epsilon_{\rm g}$. In this work we have kept $\epsilon_{\rm g}$ a free parameter and checked whether $(\epsilon_{\rm g} + 0.5)$ mod 1 is consistent with the position of the g-dominated mode in the period-\'echelle diagram (red dashed lines in panel E of Fig.~\ref{KIC010123207}). This is indeed the case. We use a Kolmogorov-Smirnov test to check whether the distribution of $\epsilon_{\rm g}$ that we find is consistent with the distribution of $\epsilon_{\rm g}$ fixed to 0. For the observational data (black dots in Fig.~\ref{dPvsq_epsg}) the probability of the cumulative distributions being the same is $2.6 \cdot 10^{-8}$. For the models the probability of the cumulative distributions being the same is $5.3 \cdot 10^{-8}$. In case of the combined sample the consistency is vanishingly small with a probability of $1.8 \cdot 10^{-12}$. As a final check, we performed the analysis as described in Section 2 with $\epsilon_{\rm g}$ fixed at 0. In these cases  $(\epsilon_{\rm g} + 0.5)$ mod 1 is in a number of cases not consistent with the location of the g-dominated modes in the period-\'echelle diagram. Hence, we conclude that $\epsilon_{\rm g}$ should not be kept fixed. 

It is noteworthy that for the RGB models 1 and 2 the value of $q$ decreased towards the theoretical value when keeping $\epsilon_{\rm g}$ fixed. As the period-\'echelle diagram did not reflect the value of $\epsilon_{\rm g}$ we do not trust these results. It does however indicate a correlation between $\epsilon_{\rm g}$ and $q$. Indeed, for the stars and models considered here we find a linear correlation between $q$ and $\epsilon_{\rm g}$ with a Pearson r coefficient of 0.34. We note that it may be that it is this correlation that is (partly) responsible for the correlation evident in Fig.~\ref{qcomp}.

The results in Fig.~\ref{dPvsq_epsg} show that there is a preference for $\epsilon_{\rm g}$ to be larger than 0.3 for RGB stars. Our sample of stars is too small to judge whether this is a coincidence, due to selection effects or real.

\begin{figure}
\includegraphics[width=\linewidth]{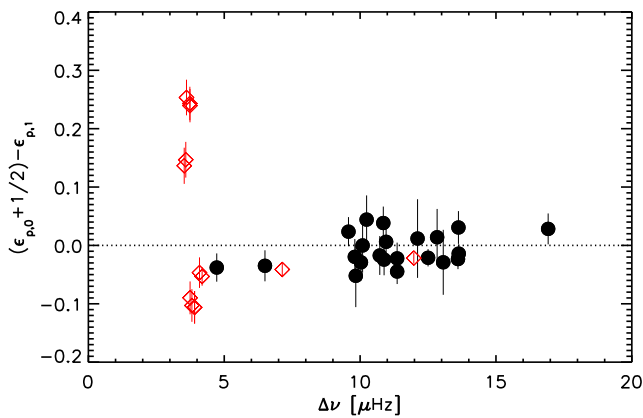}
\caption{Values of $(\epsilon_{\rm p\,0} + 0.5)-\epsilon_{\rm p\,1}$ vs. $\Delta\nu$ for observed stars (black dots) and models (red diamonds).}
\label{epsp1}
\end{figure}

\subsection{Offset $\epsilon_{\rm p\,1}$}
As discussed previously, the frequencies at which the radial and dipole pressure modes are observed are described with the help of two offsets, $\epsilon_{\rm p\,0}$ for the radial modes and $\epsilon_{\rm p\,1}$ for the dipole modes. Here we investigate the assumption that $\epsilon_{p\,l}=\epsilon_{p\,0}+l/2$ (Section 2). We show in Fig.~\ref{epsp1} that indeed most of the $(\epsilon_{\rm p\,0} + 0.5)-\epsilon_{\rm p\,1}$ values (that is equivalent with $\delta\nu_{01}/\Delta\nu$) for the RGB stars cluster around zero. We note a decrease of $(\epsilon_{\rm p\,0} + 0.5)-\epsilon_{\rm p\,1}$ for CHeB models towards lower values of $\Delta\nu$ similar to what was presented by  \citet{corsaro2012} for $\delta\nu_{01}/\Delta\nu$ for CHeB stars. The main discrepancies lie in the five models that have values for $(\epsilon_{\rm p\,0} + 0.5)-\epsilon_{\rm p\,1}$ above 0.1. These are CHeB models 0, 1, 3, 4 and 7.  For these models we checked the frequencies and the values for $\epsilon_{\rm p\,0}$ and $\epsilon_{\rm p\,1}$ obtained by our method do reflect the locations of the radial modes and p-dominated dipole modes, respectively. Therefore, it may be that there is something with the structure of the model. We find that the models that have positive values of $(\epsilon_{\rm p\,0} + 0.5)-\epsilon_{\rm p\,1}$ are all the original models or have mode inertias that have not shifted compared to the original models.

\subsection{Cavity boundaries}
\begin{figure}
\includegraphics[width=\linewidth]{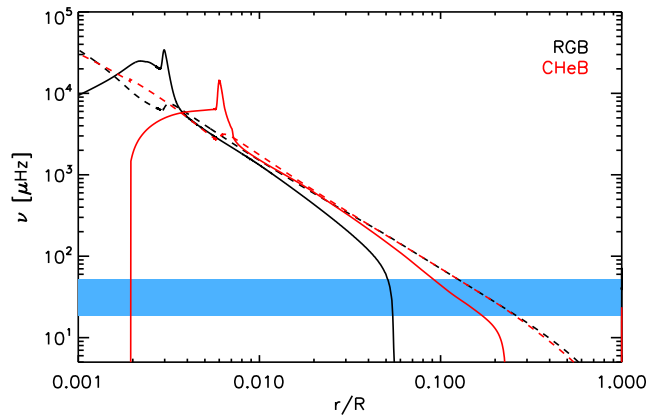}
\caption{Propagation diagram of a RGB (black) and CHeB model (red) with the Brunt-V\"ais\"al\"a frequencies indicated in the solid lines and the Lamb frequencies with the dashed lines. The region in which we can expect oscillations to be observed is indicated with the blue bar.}
\label{propD1}
\end{figure}

Recently, \citet{mosser2017} suggest that the local density contrast of the core $\beta_N = - \frac{\mathrm{d} \ln N}{ \mathrm{d} \ln r}$ and the envelope $\beta_S =  - \frac{\mathrm{d} \ln S_1}{ \mathrm{d} \ln r}$ are approximately equal:
\begin{equation}
- \frac{\mathrm{d} \ln N}{ \mathrm{d} \ln r}=\beta_N \simeq \beta \simeq \beta_S = - \frac{\mathrm{d} \ln S_1}{ \mathrm{d} \ln r}.
\label{eq:beta}
\end{equation}
This suggestion is based on the fact that the Brunt-V\"ais\"al\"a frequency and the Lamb frequency show similar radial variations for the frequencies probed in the region between the hydrogen-burning shell and the base of the convective envelope where the evanescent zone is located. This assumption is based on the analysis by \citet{takata2016a}, which is inspired by models with $\Delta\nu > 20$~$\mu$Hz. For our analysis of more evolved (RGB stars with $\Delta\nu < 15$~$\mu$Hz) the propagation diagram (Fig.~\ref{propD1}) shows that this seems a valid approximation for CHeB stars, but not for the RGB stars analysed in this work that are more evolved than the ones addressed by \citet{takata2016a}.

\begin{figure}
%\centering
%\begin{minipage}{0.48\linewidth}
\includegraphics[width=\linewidth]{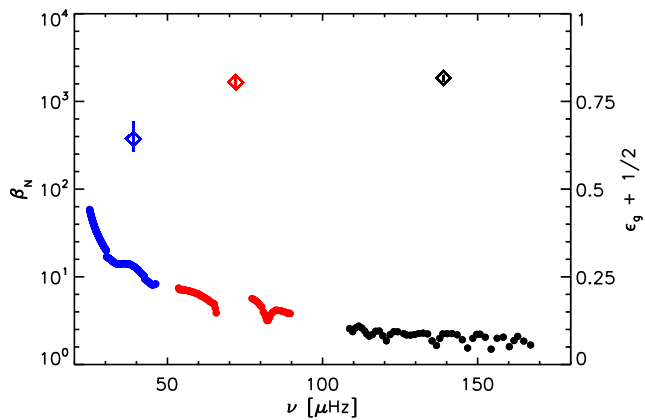}
%\end{minipage}
%\begin{minipage}{0.48\linewidth}
%\includegraphics[width=\linewidth]{RGBprop_diagram}
%\end{minipage}
%\begin{minipage}{0.32\linewidth}
%\includegraphics[width=\linewidth]{dlnNdlnSallfreqsmiddle}
%\end{minipage}
\caption{The local density contrast of the core $\beta_N$ vs frequency for the three RGB models. The diamonds indicate the value of $\epsilon_{\rm g}$ obtained for each model, considering all dipole frequencies in a 5$\Delta\nu$ wide frequency range, as per the right-hand axis. The colour-coding is the same as in Fig.~\ref{qvsqobs}.}
\label{beta}
\end{figure}

Further investigation of $\beta_N$ shows that these values increase with decreasing value of frequency for RGB stars (Fig.~\ref{beta}). In this figure we removed oscillation modes for which the computation of $\kappa$ (Eq.~\ref{eq:kappa}) was hampered by the spike in the Brunt-V\"ais\"al\"a frequency due to the discontinuity in the mean molecular weight at the deepest extent of the convection zone. 
We note that $\beta_N \cong 1$ for CHeB stars and does not show significant variations with frequency.

\citet{takata2016a,takata2016b} and references therein, show that $\epsilon_{\rm g}$ depends on the phase lags introduced at both the inner and outer turning points of the wave as well as the reflection coefficient of the edges. Fig.~\ref{beta} indicates that $\beta_N$ shows trends with $\nu$ (and with $q$). We expect that this could induce a different reflection coefficient and hence could show a correlation with $\epsilon_{\rm g}$. We indeed see a decrease in $\epsilon_{\rm g}$ with the evolution along the RGB (that is with models with decreasing frequencies) indicated with the diamonds and the right axis in the left panel of Fig.~\ref{beta}. This could indicate that a larger density contrast at the edge of the g-mode cavity would cause a lower offset $\epsilon_{\rm g}$. More models are required to confirm this trend.

For $\beta_S$ \citep{takata2016b} predict an upper limit for RGB stars of 1.5. We find consistent results for our models with a roughly constant value of $\beta_S \cong 1.35$, with an increasing value towards lower frequency. 

\subsection{Impact of different mode sets and correlations}

In some cases different mode sets arise from independent analysis of the stars, in other cases we  choose different mode sets from stellar models using different criteria to select the modes.

For the different mode sets of the stars KIC~9145955 and KIC~10200377 we see substantial overlap in the detected modes and their frequencies (see Fig.~\ref{freqs}), which results in derived values for $\Delta\Pi$, $q$, $\epsilon_{\rm g}$, $\epsilon_{\rm p\,1}$ and $\Delta\nu$ that differ outside the quoted uncertainties. As $\Delta\nu$ is obtained from the radial modes and is kept fixed during the remainder of the procedure, the difference in this parameter may have impact on the other results. So, we see that the results are frequency dependent and that formal uncertainties as quoted here do not account for that.

Furthermore, for both the models and the observed stars we find clear correlations between the obtained $\Delta\Pi$ and $\epsilon_{\rm g}$. All different mode sets of a particular model show a trend in which $(\epsilon_{\rm g} +0.5)$~mod~1 decreases with increasing $\Delta\Pi$ as expected from Eq.~\ref{eq:mixfreq}. For the CHeB models we additionally see that 1) the wider frequency range always leads to lower values of $(\epsilon_{\rm g} +0.5)$~mod~1 and a higher value of $\Delta\Pi$ and 2) that this is the case irrespective of the selections of the frequencies in this range (at least for the two sets we investigated here). For the RGB models the correlation with the different datasets with different frequency ranges is not so clear.

The trends that are present in our results are similar to the ones between $\Delta\nu$ and $\epsilon_{\rm p}$ for the acoustic modes. For the acoustic modes these differences are related to the fact that $\Delta\nu$ is a function of frequency due to stellar structure changes on long and short scales. From theory it is also expected that the period spacing is frequency dependent, which is what appears in our current results. This together with the fact that we find systematic differences between $\Delta\Pi$ obtained from frequencies and from the integral of the Brunt-V\"ais\"al\"a frequency is a direct indication that a comparison of $\Delta\Pi$, as well as other parameters, obtained from the same sets of frequencies in both observations and models is essential for detailed comparisons.

\section{Conclusions}
In this work we investigated the use of the formalism by \citet{jiang2014} for red-giant branch and core helium burning stars to obtain values for $\Delta\Pi$, $q$, $\epsilon_{\rm g}$ and $\epsilon_{\rm p\,1}$ from individual frequencies. This formalism provides a global solution based on all dipole modes with the same azimuthal order and can be applied to all azimuthal orders for which the results are combined in the current work.  

The fact that the radial order is explicitly included in this formalism reduces problems with alias results that are present in other methods and provides the possibility to constrain $\epsilon_{\rm g}$. On the other hand we find that for cases with weak coupling e.g. KIC~5866737 the lower number of frequencies and the higher radial order provide challenges to the method resulting in reduced reliability of the results.
\newline
\noindent The current results indicate:
\begin{itemize}
\item that for RGB stars $(\epsilon_{\rm g}+0.5)$~mod~1 can be constrained and is in all cases analysed here between 0.3 and 1;
\item that the local density contrast at the edge of the g-mode cavity ($\beta_N$) does follow a trend with $\epsilon_{\rm g}$. This needs further analysis;
\item that there is  systematic overestimation of $\Delta\Pi$ for CHeB stars when computed from frequencies compared to the asymptotic value computed from the integral of the Brunt-V\"ais\"al\"a frequency, as already mentioned in the literature; 
\item that $\Delta\Pi$ and $(\epsilon_{\rm g}+0.5)$~mod~1 depend on the mode set from which they are determined, where for CHeB models mode sets covering a wider frequency range provide higher values for $\Delta\Pi$ and lower values for  $(\epsilon_{\rm g}+0.5)$~mod~1. These trends are typically not included in the quoted uncertainties. To mitigate this when performing a model comparison we deem it vital to treat data and models in the same way for a meaningful result;
\item the values for $\epsilon_{\rm p\,1}$ for the CHeB models may indicate that additional physics has to be included in the models as presented by \citet{constantino2015}.
\end{itemize}

From the models we find a linear correlation between the relative width of the evanescent zone normalised by its location ($\Delta r / r_{\rm evanescent}$) and the natural logarithm of the coupling factor $q$.

To further explore these conclusions and investigate the differences and trends in more depth, larger sets of observed stars and systematically chosen models for which individual frequencies are available need to be investigated. Developments of methods to extract individual frequencies from the power spectra of timeseries data are being developed \citep[e.g.][Garcia Saravia Ortiz de Montellano et al. in preparation]{corsaro2015} and the frequencies of larger sets of observed stars are expected in the near future.

\begin{acknowledgements}
We thank Abishek Datta, Anwesh Mazumdar and Thomas Constantino for kindly providing us with the models for RGB stars \citep{datta2015} and CHeB stars \citep{constantino2015}. Additionally, we thank the anonymous referee, Joergen Christensen-Dalsgaard, Enrico Corsaro, Benoit Mosser and Masao Takata for comments on earlier drafts that improved the manuscript considerably.
The research leading to the presented results has received funding from the
European Research Council under the European Community's Seventh Framework
Programme (FP7/2007-2013) / ERC grant agreement no 338251 (StellarAges).
YE  acknowledges the support of the UK Science and Technology Facilities Council (STFC).
\end{acknowledgements}

\bibliographystyle{aa}
\bibliography{Pspacing}

\begin{appendix}
\section*{ }
\addtocounter{section}{1}
\begin{figure}
\centering
\includegraphics[width=\linewidth]{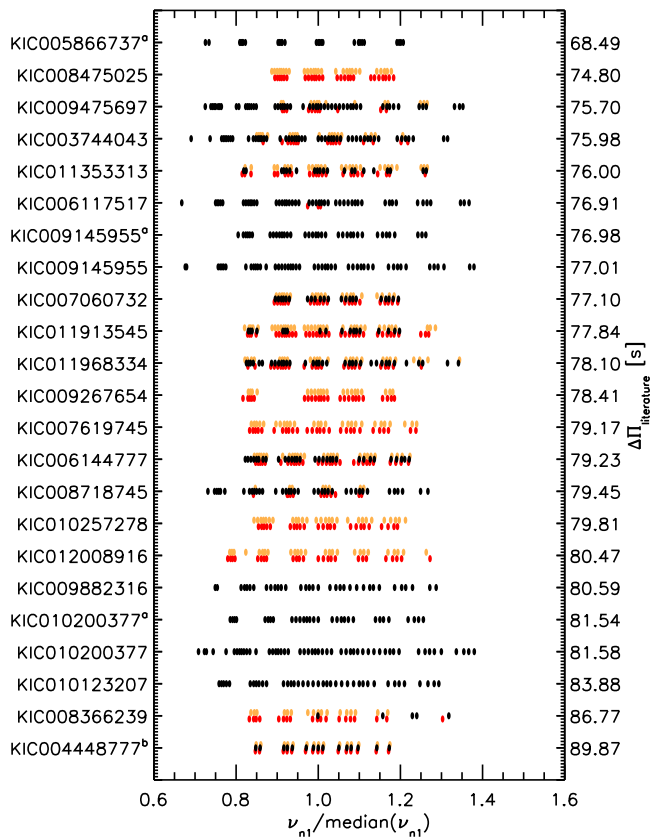}
\caption{The dipole frequencies of all RGB stars used in the present analysis normalised by the median dipole frequency of the star. Modes with azimuthal orders $-1,0,1$ are shown (with a slight offset to one another) in red, black and orange, respectively. Stars names with a superscript `a' and `b' indicate that the data are taken from \citet{datta2015} and \citet{dimauro2016}, respectively. All other data are taken from \citet{corsaro2015}. On the left the period spacings presented in these references are shown (including the updates by Corsaro, see text). The stars are ordered following the red-giant branch with the least evolved star at the bottom and the most evolved star at the top.}
\label{freqs}
\end{figure}
\end{appendix}

%\appendix

%\section{Results for all stars}
%Here we present the results for the individual stars analysed in this work, both in Tables and Figures.

 \end{document}